\documentclass[a4paper]{jpconf}
\usepackage{graphicx}
\usepackage{citesort}
\begin{document}
\title{Proton induced activity in graphite - comparison between measurement and simulation}

\author{Daniela Kiselev, Ryan Bergmann, Dorothea Schumann, Vadim Talanov and Michael Wohlmuther}

\address{Paul Scherrer Institut, 5232 Villigen PSI, Switzerland}

\ead{Daniela.Kiselev@psi.ch}

\begin{abstract}
The Paul Scherrer Institut (PSI) operates the Meson production target stations E and M with 590 MeV protons at currents of up to 2.4 mA.  Both targets consist of polycrystalline graphite and rotate with 1 Hz due to the high power deposition (40 kW at 2 mA) in Target E. The graphite wheel is regularly exchanged and disposed as radioactive waste after a maximum of 3 to 4 years in operation, which corresponds to about 30 to 40 Ah of proton fluence. For disposal, the nuclide inventory of the long-lived isotopes (T$_{1/2}$ $>$ 60 d) has to be calculated and reported to the authorities. Measurements of gamma emitters, as well as $^3$H, $^{10}$Be and $^{14}$C, were carried out using different techniques. The measured specific activities are compared to Monte Carlo particle transport simulations performed with MCNPX2.7.0 using the BERTINI-DRESNER-RAL (default model in MCNPX2.7.0) and INCL4.6/ABLA07 as nuclear reaction models.   
\end{abstract}

\section{Motivation and introduction}
When radioactive material has to be disposed of, the authorities in Switzerland require a complete nuclide inventory for long-lived isotopes (T$_{1/2}$ $>$ 60 d). A complete nuclide inventory can only be achieved with reasonable effort via simulation, and for important nuclei, these calculations have to be validated. Often measuring important isotopes is not straight-forward and requires a chemical treatment beforehand.  The authorities are not only interested in  the radioisotopes with the highest activities or dose rates, but also in isotopes which are particularly problematic for the environment, such as $\alpha$-emitters, volatile or outgassing substances, and elements known for migration or chemical reactions in final repositories.

Under irradiation of 590 MeV protons, mainly $^3$H, $^7$Be, $^{10}$Be and $^{14}$C are produced in pure graphite. $^7$Be has a short life time of only 53 days, and it is usually already completely decayed when the graphite targets are ready for disposal. However, the activity of $^7$Be determines the high dose rate of a few Sv/h at the graphite wheel a few days after operation in-beam. Contrary to it, $^{10}$Be and $^{14}$C are very long-lived. Carbon in an oxidized gaseous state is a candidate for migration out of the final repository into the biosphere and geosphere. Therefore,  when disposing radioactive graphite, special measures are taken which are known from decommissioning of nuclear reactors. 
Tritium with a half-life of 12.3 a is produced in large quantities due to spallation (fragmentation of nucleus). This isotope is known for later outgassing from components. Since graphite is a porous material, it is expected that more tritium is released in graphite than from widely used metals like steel and copper. However, due to its low energy $\beta$-particle, tritium is not very harmful. It is also released in large quantities from nuclear power plants. 

In addition, graphite used as moderator sticks in nuclear reactors is usually accompanied by large quantities of $^{36}$Cl, a long-lived radioisotope, which is also a candidate for migration and outgassing. 

In this report, a comparison is made of the calculated and measured specific activities of very pure graphite irradiated by a 590 MeV proton beam.  The comparison is done on the following radioisotopes: $^3$H, $^7$Be, $^{10}$Be, $^{14}$C, $^{22}$Na, $^{26}$Al, $^{44}$Ti, $^{54}$Mn. Isotopes with mass equal to and lower than carbon are produced directly from carbon, whereas the heavier isotopes are produced from impurities.  Besides the uncertainties in the impurities, this simple system is well suited for benchmarking.  In addition, the long standing puzzle to reproduce the large  activity of $^{14}$C measured in \cite{Argentini00,Argentini02} by LAHET \cite{lahet} or later MCNPX2.5.0 \cite{mcnpx250}  calculations \cite{Kiselev06} makes this system particularly interesting. The experimental values in \cite{Argentini00} were later reproduced by \cite{Schumann12}.  Two nuclear reaction models are used for comparison: The default model in MCNPX2.5.0 \cite{mcnpx250} and MCNPX2.7.0 \cite{mcnpx270}, the BERTINI-DRESNER-RAL \cite{Bertini63,Bertini69,Dresner81,Atchison07}, as well as the improved INCL4.6/ABLA07 \cite{Boudard13,Kelic08}.

\section{Meson production target station E}
  At PSI there are two meson production target stations, called M and E. Both consist of a graphite target and deliver not only mesons, i.e. mainly pions, which later decay to muons. Pions and muons are used for research in particle physics and for muon spin-resonance applications ($\mu$SR).  
  
The meson target stations are driven by 590 MeV protons with beam currents of up to 2.4 mA from the High Intensity Proton Accelerator (HIPA), which delivers the most intense continuous proton beam in the world. The protons are produced in a compact electron cyclotron resonance (ECR) source \cite{Baumgarten11} and gain 870 keV by a Crockcroft-Walton generator. Afterwards they are further accelerated in two cyclotrons called ``Injector2'' and ``Ring''. The protons leave the Injector2 with an energy of 72 MeV and are transferred to the Ring, where they are accelerated to their final energy of 590 MeV. The beam is transferred to the first meson production target station, Target M, with an effective graphite thickness of 5 mm, or optionally kicked to the ultra-cold neutron source (UCN) for a few seconds. About 18 m later, the beam passes through the meson production target station, Target E, with an variable thickness of 40 mm or 60 mm.  In Target E region, the protons lose 15 MeV of their initial energy and about 30 \% of their intensity. The losses are caused mainly due to multiple scattering in the graphite of Target E, which also leads to additional divergence of the beam. For further transport, the beam is shaped by a subsequent collimator system, which is responsible of 2/3 of the beam losses. The beam loses at the Target M region are only 1 \%.  After the meson production targets, the beam is bent downwards then upwards to irradiate the spallation neutron source SINQ from beneath. At SINQ, thermal and cold neutrons are produced for material structure research and radiography from a zircalloy-clad lead target.  If SINQ is not ready to receive beam, protons are instead steered to the beam dump after the meson production targets, keeping Target E and Target M in operation.

\begin{figure}[ht]
\includegraphics[width=14pc]{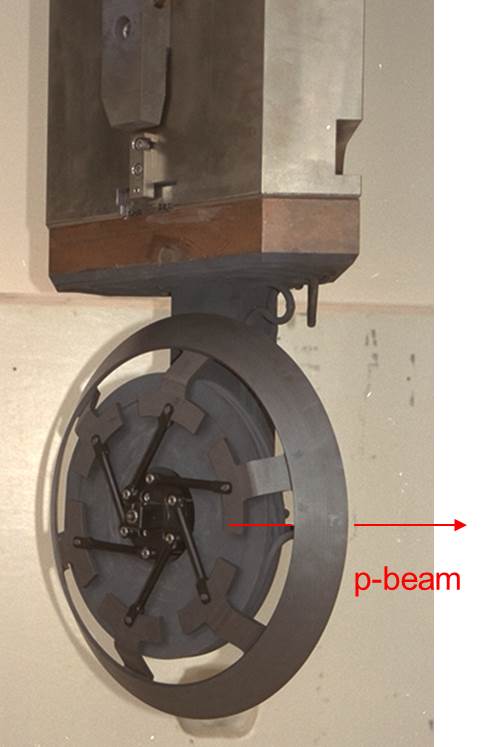}\hspace{2pc}%
\begin{minipage}[b]{14pc}\caption{\label{einschub}Insert of target E. The direction of the beam is also indicated.}
\end{minipage}
\end{figure}

Only samples from Target E are investigated in this report. The insert of Target E is shown in figure \ref{einschub}. Two inserts exist since the bearing might have to be changed in one operation period. Two challenges for Target E are the high radiation levels as well as the high power deposition. The 590 MeV protons of 2.4 mA deposit about 50 kW in the 40 mm thick graphite of Target E and even more in the 60 mm target. To reduce the beam spread, a light material was chosen for the target material. Beryllium could not sustain the radiation dose, but polycrystalline graphite could. To dissipate the heat, the target was designed as a wheel rotating at 1 Hz. In the vicinity of the target, water-cooled copper removes the radiated heat. The graphite is heated to about 1500 $^o$ C at the beam spot. As graphite is an anisotropic material by nature, the thermal expansion coefficients are dependent on the lattice axis. This leads to extra thermal stress and finally to deformation or even cracks. Therefore, polycrystalline graphite, which consists of a large number of small single crystallites irregularly arranged in space, is used. This results in almost isotropic physical properties.   

\subsection{Description of sample Targets}
The three targets, called E89, E83 and E92, which were used for this analysis were all made from SGL-CARBON graphite but from different grades. The target E79 uses EK94, E83 was produced from R6400, and E92 is made of R6510. The later grades have improved isotropic properties. They have smaller grains which lead to less porosity, slightly larger density, and better radiation resistance. EK94 has a density of 1.73 g/cm$^3$, R6400 1.75 g/cm$^3$ and R6510 1.83 g/cm$^3$. For all three materials, a material analysis using ICP-OES (Induced Coupled Plasma Optical Emission Spectroscopy) was performed. A tendency toward smaller impurities in the newer grade R6510 can be observed with respect to EK94, the oldest grade. However, a material analysis performed with INAA (Instrumental Neutron Activation Analysis) showed much smaller impurities for most trace elements.  Due to these discrepancies, the impurity concentrations were set so the calculated activities matched the measured activities. 

In addition to the above mentioned material investigations, GDMS (Glow Discharge Mass Spectrometry) was performed to measure the gaseous elements N and Cl. Nitrogen was thought to contribute significantly to the production of $^{14}$C via the $^{14}$N(n,p)$^{14}$C reaction. A content of 15 ppm N was measured in EK94. For nitrogen contents larger than 1 ppm, $^{14}$C is predominately produced via this reaction, in nuclear reactors \cite{Heikola14}. In spite of the missing  $^{14}$C activity in the calculation as mentioned in the introduction, the effect of adding nitrogen to pure carbon was investigated.  

In reactor graphite $^{36}$Cl is known ro be produced in large quantities from chlorine.  As $^{36}$Cl is a long-lived gaseous radioisotopes which is able to migrate out of the final repository, special care is taken when disposing reactor graphite.  Furthermore, chlorine outgasses more readily from graphite than from metals due to its porous structure.  Therefore chlorine was measured in the material composition and later the resulting $^{36}$Cl was calculated (s. section \ref{sum}).

The results of the material analyses for the isotopes of interest are shown in table \ref{material}. The results from ICP-OES for the three graphite grades are presented as ranges.   

\begin{table}
\caption{\label{material}Material analysis from ICP-OES and GDMS.}
\begin{center}
\begin{tabular}{ll}
\br
element&content [ppm]\\
\mr
N   &15\\
Na & 13 - 77 \\
Al  & 6 - 25 \\
Cl  & 1.1 \\
Ti  & 0.4 - 1.0 \\
Mn & $\approx$ 0.1 \\
\br
\end{tabular}
\end{center}
\end{table}

The Targets E79 and E83 were irradiated for only a few months and disposed as radioactive waste some years later.  Samples were taken at disposal time, and the activities were measured shortly after.  The measurement of the activity was performed again after 7 and 10 years when short-lived radioisotopes like $^{7}$Be could not be detected anymore. More precise values of the sample irradiation and cooling period from both targets can be found in tables \ref{irrad79} and  \ref{irrad83}.

\begin{table}[ht]
\begin{center}
\begin{minipage}{7cm}
\caption{\label{irrad79}Irradiation history of E79.}
\begin{tabular}{ll}
\br
period [years]&charge [Ah]\\
\mr 
 0.6 & 6.6 \\
 9.9 & measurement \\
 \br
\end{tabular}
\end{minipage}
\begin{minipage}{7cm}
\caption{\label{irrad83}Irradiation history of E83.}
\begin{tabular}{ll}
\br
period [years]&charge [Ah]\\
\mr 
 0.2 & 1.8 \\
 7.3 & measurement \\
 \br
\end{tabular}
\end{minipage}
\end{center}
\end{table}
 
Target E92 was irradiated to a larger proton fluence of 28.8 Ah. It was irradiated in 4 periods with proton fluences ranging between 6 and 8 Ah with cooling periods of a few months in between. The measurement of the activities was performed 4 years after the last irradiation. In this case, $^{7}$Be could still be detected. The irradiation history of E92 can be found in table \ref{irrad92}.

\begin{table}[ht]
\begin{center}
\caption{\label{irrad92}Irradiation history of E92.}
\begin{tabular}{ll}
\br
period [years]&charge [Ah]\\
\mr 
 0.5 & 6.6 \\
 0.3 & 0  \\
 0.5 & 6.3 \\
 0.5 & 0 \\
 0.7 & 7.4 \\
 0.3 & 0 \\
 0.7 & 8.5 \\
 3.8 & measurement \\
 \br
\end{tabular}
\end{center}
\end{table}

\section{Experimental determination of specific activities in graphite}
In the graphite samples the following radioisotopes were detected:  $^3$H, $^7$Be, $^{10}$Be, $^{14}$C, $^{22}$Na, $^{26}$Al, $^{44}$Ti, $^{54}$Mn. The gamma emitters  $^7$Be, $^{22}$Na, $^{26}$Al, $^{44}$Ti, $^{54}$Mn were detected by a High Purity Germanium detector (HPGe) from ORTEC and analyzed by the software GENIE2000 of Canberra. $^7$Be could only be detected in the sample of E92 as the cooling times for the other two targets were too long (s. tables \ref{irrad79}, \ref{irrad83}, \ref{irrad92}). 

$^{10}$Be, $^3$H and $^{14}$C were extracted by chemical treatment (wet digestion) as described in reference \cite{Schumann08}. A small amount of graphite (20 mg) was boiled in a mixture of acids H$_2$SO$_4$/HNO$_3$/HClO$_4$ and washed out with nitrogen. In the flaks filled with HCL tritium remains in the dissolving solution. In the second flask the carbon dioxide absorbs in the NaOH. $^{3}$H and
$^{14}$C measurements were carried out using liquid scintillation counting (LSC). To determine the absolute activity the yield during the chemical treatment have to be known. Therefore, inactive graphite spiked with an known amount of isotope standard underwent the same chemical treatment as the samples. Afterwards the products were measured under identical
conditions. The chemical yields obtained were about 95 \% for $^{14}$C and about 85 \% for $^{3}$H \cite{Schumann12}. The uncertainty of the $^{14}$C measurement is 15 \% and the detection limit is 500 Bq/g. 

The $^{10}$Be content was measured by Multi-Collector Inductively-Coupled- Plamsa Mass Spectrometry (MC-ICP-MS).

\section{Calculation}
The calculation of the nuclide inventory was done with the Monte Carlo particle transport code MCNPX2.7.0 \cite{mcnpx270} using the rnucs patch \cite{Gallmeier15} to account for spallation-induced residuals. In this chapter, the method as well as the applied reaction models will be described shortly.

MCNPX2.7.0 can transport a large variety of particles: protons, neutrons, pions as well as ions like deuterons, tritons up to heavy nuclei. As we will see in the following, light ions are important to include as they can make significant contributions to the activation. For the nuclear reactions of all particles, physics models have to be used when no cross section tables are available or selected. One class of nuclear reaction models are based on the  intranuclear cascade (INC) mechanism followed by evaporation or fission. These models are usually applicable from about 150 MeV to a few GeV. However, often they are used for lower energies from 20 MeV on, where some cross section tables stop working.  In INC, the energy of the primary particle is dissipated in the nucleus to several nucleons by nucleon-nucleon collisions until equilibrium is reached. In the beginning of this phase, the so called pre-equilibrium, particles, which receive a large energy transfer by direct interaction, can leave the nucleus with large kinetic energy in forward direction (relative to the incoming particle).  Later, the nucleus is in an excited state with an often nonzero angular momentum. Both energy and angular momentum can be released by either emission of light particles like photons, neutrons, deuterons etc. (evaporation) or by splitting the nucleus in two (fission). Large particles like Mn can be emitted during the evaporation process as well, but with lower probability. As the particles are emitted at relatively low energy in the evaporation phase, they are isotropically distributed. Fission occurs mainly for heavier nuclei like Pb as these are more affected by surface vibrations. The evaporation and fission is often treated by separate models, to which the INC code is coupled.  The INC code BERTINI  \cite{Bertini63,Bertini69} is often coupled to the evaporation code DRESNER \cite{Dresner81} and the fission model RAL \cite{Atchison07}.  The DRESNER code is based on Weisskopf´s statistical model. This combination has been the default choice in MCNPX for neutrons and protons of energy less than 3.5 GeV for years. As BERTINI can handle only protons and neutrons, reactions containing deuterons, tritons and $\alpha$-particles up to 1 GeV are handed over to ISABEL \cite{isabel}. ISABEL can also replace BERTINI completely.  

MCNPX2.7.0 contains an outdated version of INCL/ABLA (INCL4.2/ABLAV3p), which has a couple of well-known shortcomings. Therefore, in the modified version that was used, INCL was upgraded to INCL4.6 \cite{Boudard13} and ABLA to ABLA07 \cite{Kelic08}. Up to now this model is not yet in one of the distributed versions of MCNPX or MCNP.  With the help of J.-Chr. Davide, INCL4.6/ABLA07 was implemented in the version of MCNPX2.7.0 in such a way that it could be run on multiple processors.  As the name implies, the INCL models are also based on INC. The new INCL4.6 model handles not only neutrons, protons, pions as in INCL4.2, but also deuterons, tritons, $^3$He, $\alpha$-particles, and light ions up to mass number A = 8. These particles cannot only induce nuclear reactions, but can also be emitted in all phases of the physics model. The light ions are formed by coalescence in phase space from neutrons and protons.  ABLA07 itself is able to generate neutrons, protons, deuterons, tritons, $^3$He, $\alpha$-particles, and intermediate mass fragments (IMF) by break-up, fission or evaporation. In addition, when using INCL4.6/ABLA07, metastable isotopes can now be handled in MCNPX2.7.0 according to the PHTLIB file as it is done for the default cross section choice.

In the following simulations, the two above mentioned models are used for all particle interactions except for neutrons with energies less than 20 MeV.  Below 20 MeV, tabulated cross sections from the ENDF/B-VII \cite{endfb} dataset or similar versions are applied. For later evaluation of the nuclide inventory, the neutron flux spectra up to 20 MeV are written to the output file of MCNPX for the cells of interest. For the other interactions controlled by the physics models, a lot of information about each reaction is usually written into a binary file called ``histp''. This file can become several GB in size. Since the calculation of the nuclide inventory requires only the production rates, this information is directly stored in the output file of MCNPX and the histp-file can be dismissed.  This information is made available in the output file by using the rnucs-card \cite{Gallmeier15}.  With the help of a PERL script (“activation script” \cite{Gallmeier08}), all information including the irradiation history and cooling times is passed to build-up and decay codes like CINDER1.05 \cite{Wilson95} or FISPACT10 \cite{Sublet10} which can use the cross section library EAF-2010 \cite{Sublet10}. In the following evaluation, the results using FISPACT10/EAF-2010 are presented. However, a check with CINDER1.05, which comes with its own library, did not lead to significant different results.  

Most simulation runs were performed for the sample from Target E92:  11 runs using BERTINI-DRESNER-RAL and 36 runs using INCL4.6/ABLA07.  The material composition and the transported particles were changed between the runs (see below). Up to 10$^8$ primary protons were needed to ensure a statistically significant result. When reactions with impurities were not being sampled sufficiently due to their low concentrations, the option ``impure'' of the rnucs-card became important. This ensures that all isotopes of the material composition undergo reactions with the same probability.  Concentration is taken into account by applying the proper weight to the tally score to ensure a fair game is played and the result means are not biased. 

The calculations were performed on a medium-performance cluster at PSI. The number of cores used was between 24 and 72. The core-hours per primary particle depend strongly on the number of particles transported and the number of isotopes in the material composition. On 24 cores, MCNPX2.7.0 using INCL4.6/ABLA07 with the extended list of particles and pure carbon needs 6 h for 10$^8$ primary protons. BERTINI-DRESNER-RAL, known as the faster physics model, was in fact a factor 2 slower as this compilation of MCNPX2.7.0 was done without compiler optimizations. 

\subsection{Geometrical model}
\begin{figure}[ht]
\begin{center}
\includegraphics[width=14cm]{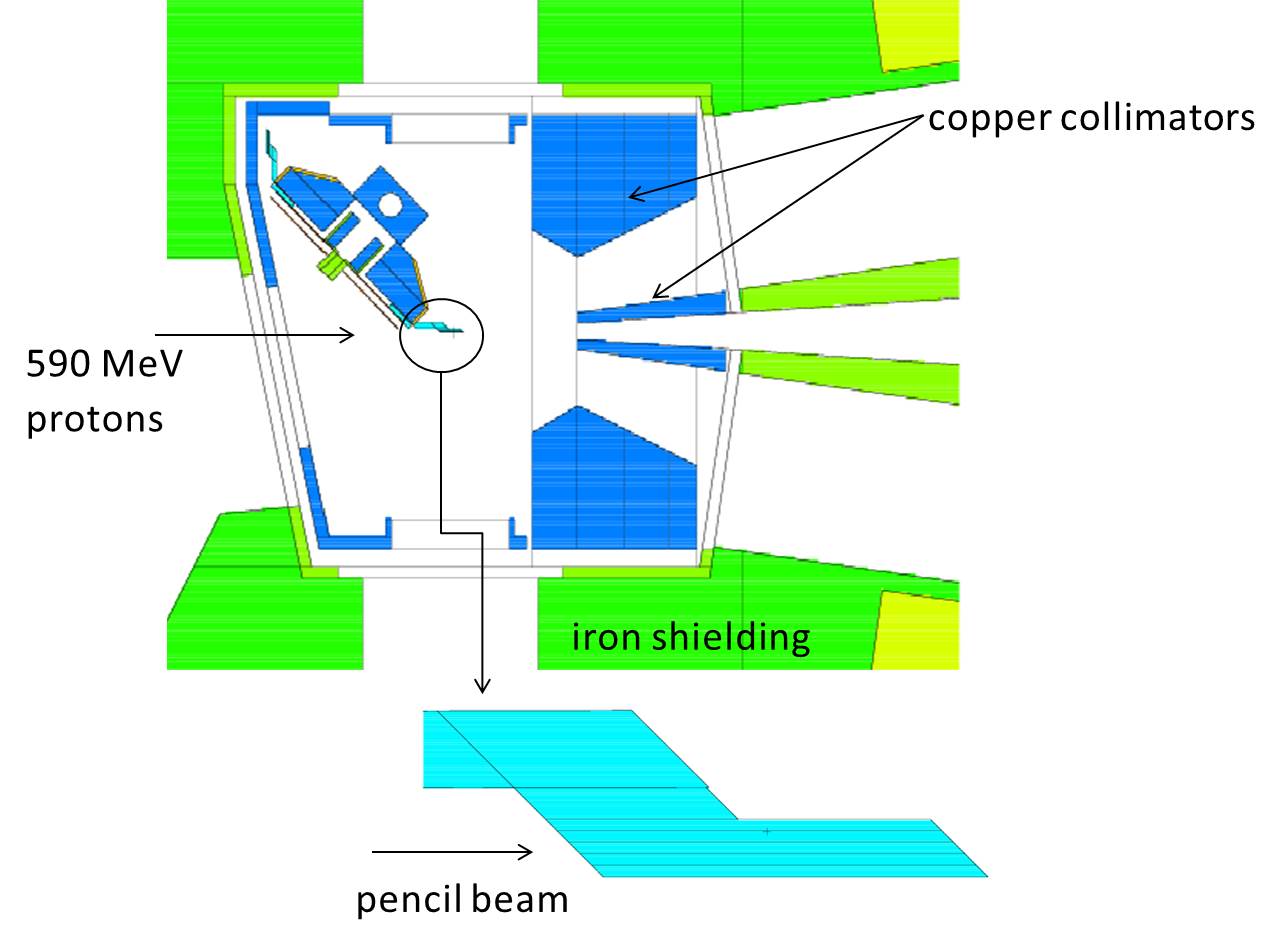}
\end{center}
\caption{\label{geom}Geometry in MCNPX  for the current calculations.}
\end{figure}

For the activation calculations, a precise geometrical model were used (see figure \ref{geom}). The reason for doing this is twofold. First, this model will be used also for other purposes like calculating the energy deposition  on the cooled water plates as well as comparing measured and calculated activity of a concrete sample located 2 m above beam height.  Second, during the former investigations regarding the nuclide inventory \cite{Kiselev06}, it was thought that the large discrepancy between calculation and measurement regarding the $^{14}$C activity is caused by backscattered neutrons, which were in the early analysis not taken into account as the geometrical model consists just of a graphite block. The backscattered neutrons would have low energy where the neutron capture cross section on $^{13}$C producing $^{14}$C is large. In figure \ref{geom}, copper parts are shown in blue. They are almost all cooled by water, which is not taken into account in the simulation. Steel parts are shown in green. The shielding around the target chamber are made from normal carbon steel. 

The rim of the graphite wheel sliced into 5 pieces in horizontal direction. For both the 40 mm and 60 mm target, the rim is 6 mm wide. The beam is centered on the rim and usually has a Gaussian distribution with standard deviations between 0.5 and 1 mm. In the vertical plane, the beam is a bit wider than in the horizontal direction.   Since the beam parameters and the position on the target vary, the simulation assumes a pencil beam at the rim center for simplicity.   

\section{Results}
In this section, the calculations and results obtained under different settings will be presented and compared to the experimental values. Except for $^{10}$Be, the experimental results were already published in \cite{Schumann12}. Most investigations were done on E92 since this is the target with the highest proton fluence and the shortest cooling time. A summary of all results will be given at the end of this report.

\subsection{$^{14}$C activity}
The main investigations were done concerning the $^{14}$C activity due to the large discrepancy of 2 orders of magnitude seen in the past.  First, using BERTINI-DRESNER-RAL in MCNPX2.7.0 this result could be confirmed. The ratio between the calculated  and  measured activities on E92 is 1/71.5. The uncertainty for the measured $^{14}$C activity is 15 \%. By default, neutrons, protons, charged and neutral pions, deuterons, tritons, $^3$He, and $\alpha$-particles were taken into account. For this and the following results a pure carbon target is assumed except when stated otherwise.  The next trial was to add 15 ppm nitrogen to the pure natural carbon since this concentration was measured in EK94. This increased the $^{14}$C activity by 35 \% - not enough to explain the large discrepancy. 

Using INCL4.6/ABLA07 on a pure carbon target, the ratio of the activities from Monte Carlo and measurement increased to 0.92, which is an enormous improvement compared to BERTINI-DRESNER-RAL, and even within the error bars of the measurement.  To figure out the reaction mechanism for the production of $^{14}$C, a simulation was performed with the ``standard'' particles neutrons (n), protons (p) and pions ($\pi$) for both models on pure carbon. This lead to very similar results for both models: using  BERTINI-DRESNER-RAL 58 Bq/g and with INCL4.6/ABLA07 80 Bq/g for $^{14}$C were obtained. The value for BERTINI-DRESNER-RAL is very similar to the one obtained with the more extended list of particles (65 Bq/g) whereas the activity calculated with INCL4.6/ABLA07 and the ``standard'' particles is 53 times smaller. This means that the main production mechanism of $^{14}$C are neither the direct production via high energetic protons nor neutron caption on $^{13}$C. Therefore, one or more of the remaining particles,  deuterons (d), tritons (t), $^3$He, $\alpha$-particles have to be responsible for the sizeable increase of the $^{14}$C production. Next, simulations with INCL4.6/ABLA07 were performed including the ``standard'' set of particles plus {\it one} particle from the remaining list. The results are shown in table \ref{INCL-mode}.

\begin{table}[ht]
\begin{center}
\caption{\label{INCL-mode}Activity of $^{14}$C using INCL4.6/ABLA07 and the given list of particles transported.}
\begin{tabular}{ll}
\br
particles & $^{14}$C activity [Bq/g]\\
\mr 
 n,p,$\pi$ & 80 \\
 + t & 2529 \\
 + d & 1100 \\
 + $\alpha$ & 592\\
 + $^3$He & 122 \\
 \br
\end{tabular}
\end{center}
\end{table}

\begin{figure}[ht]
\begin{center}
\includegraphics[width=14cm]{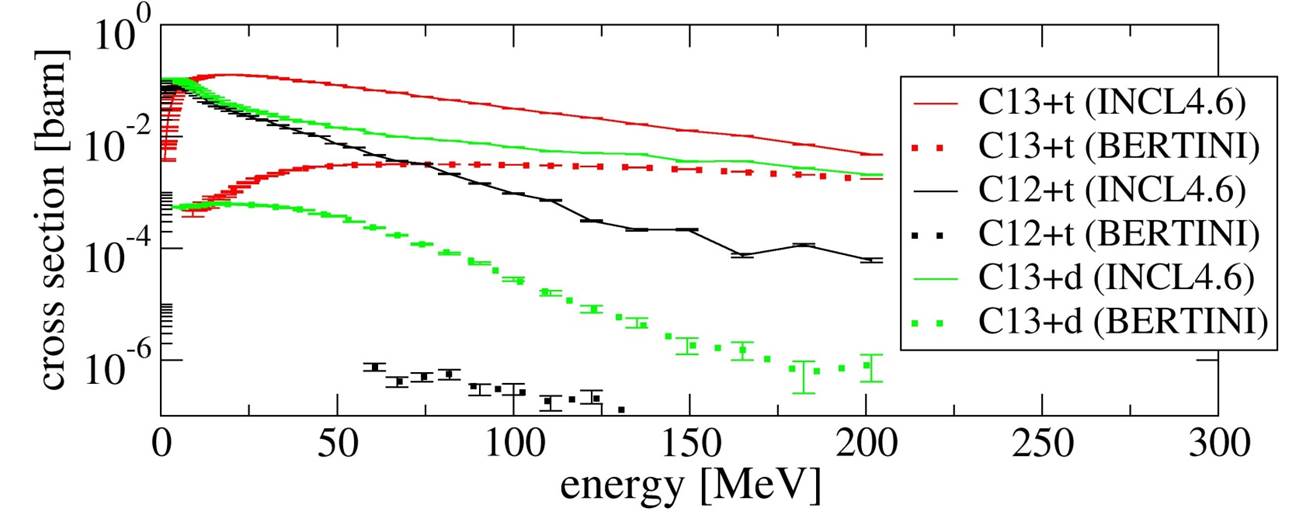}
\end{center}
\caption{\label{wq} Triton and deuteron cross sections for the production of $^{14}$C on $^{12}$C and $^{13}$C using BERTINI (dashed) and INCL4.6 (solid).}
\end{figure}

From these results, clearly tritons and deuterons produce most of the $^{14}$C. Therefore, the triton and deuteron cross sections and flux spectra using BERTINI-DRESNER-RAL as well as INCL4.6/ABLA07 were extracted from MCNPX2.7.0. The comparison of the cross sections on $^{12}$C as well as $^{13}$C are shown in figure \ref{wq}.

 In figure \ref{wq}, the cross sections extracted from BERTINI (dashed line) are always lower than the ones representing INCL4.6 (solid line).  Large differences of about three orders of magnitude can be found for tritons on $^{12}$C and deuterons on $^{13}$C.  Although the cross sections of tritons on $^{13}$C is larger than $^{12}$C, the abundance of $^{13}$C is only 1.1 \%.  The production of $^{14}$C by deuterons on $^{12}$C is negligible in both models and is not shown in figure \ref{wq}. 

\begin{figure}[ht]
\begin{center}
\includegraphics[width=14cm]{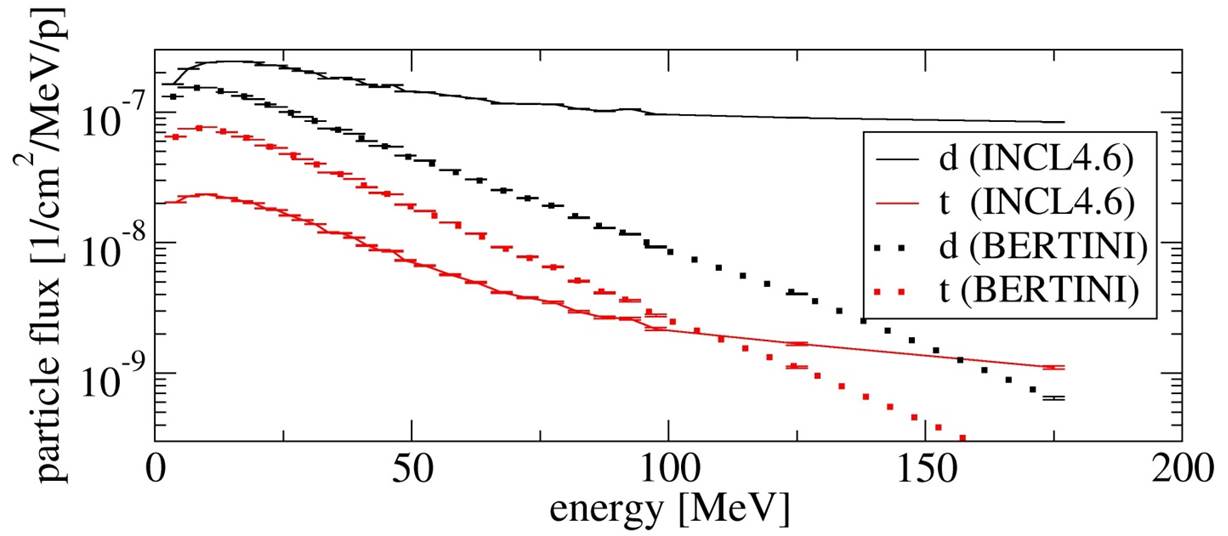}
\end{center}
\caption{\label{spec}Triton and deuteron flux spectra using BERTINI-DRESNER-RAL and INCL4.6/ABLA07.}
\end{figure}

For the production of $^{14}$C, not only the cross sections are important but also the number of incident particles and their energy distribution. The flux spectra of tritons and deuterons as a function of energy are shown in figure \ref{spec}. The results for BERTINI-DRESNER-RAL (dotted) look similar in shape for both species as well as the spectra using INCL4.6/ABLA07 (solid line). The spectra from INCL4.6/ABLA07 have an additional tail at higher energy, which becomes clearly visible for energies larger than 100 MeV. This is due to the emission of higher energetic particles during the pre-equilibrium phase. Interestingly, the triton flux from INCL4.6/ABLA07 is significantly smaller for energies less than 100 MeV than the corresponding spectrum from BERTINI-DRESNER-RAL. 

From the triton and deuteron cross sections and energy distributions, significant differences in the production of $^{14}$C are obvious, which are responsible for the increased $^{14}$C activity using INCL4.6/ABLA07 and the good agreement to the experimental data. For E79 and E83, the ratio of the $^{14}$C activity from INCL4.6/ABLA07 to the measurement are 0.50 and 1.16, respectively. The $^{14}$C activity in E83 was already a factor 2 below the detection limit, however, the measurement is in good agreement with the prediction. Averaging the results from the three samples leads to 0.86 for pure carbon. The largest effect would be a 15 ppm nitrogen contribution which would result in 30 \% more $^{14}$C and increase the ratio to 1.1. However, it is not clear, if the nitrogen content is the same in all targets as the nitrogen content was measured in EK94 only. Particularly, R6510 has a 7 \% higher density and therefore less porosity, which leads to the expectation of smaller amounts of nitrogen. In addition one has to consider that nitrogen might be completely outgassed during operation when the graphite is heated to about 1500 $^o$C. 

\subsection{Be activity}
In all three target samples, $^{10}$Be was determined experimentally. The short-lived $^{7}$Be isotope could be measured in E92 only. For E92, simulations using pure carbon were performed with both physics models. BERTINI-DRESNER-RAL agrees well with the experimental data for both Be-isotopes. The ratio Monte Carlo to measurement is 1.32  and 1.01 for $^{7}$Be and $^{10}$Be, respectively. The MCNPX2.7.0 simulation with INCL4.6/ABLA07 leads to higher ratios: 1.84 and 1.94 for $^{7}$Be and $^{10}$Be. In case of $^{7}$Be, INCL4.6/ABLA07 results in 40 \% more activity copmared to BERTINI-DRESNER-RAL. For $^{10}$Be the increase is even higher at 90 \%. 

A similar trend is observed for the other two targets. Here only INCL4.6/ABLA07 simulations were performed as the result for the calculation with BERTINI-DRESNER-RAL can be estimated from the comparison to INCL4.6/ABLA07 done for E92. For E79, INCL4.6/ABLA07 leads to 1.48 for the $^{10}$Be activity ratio, whereas for E83 the ratio is 3.02. The larger discrepancy might have also to do with the fact that E83 was irradiated to the lowest charge, and therefore the uncertainty in the measurement might be larger. For BERTINI-DRESNER-RAL, a ratio of about 1.6 can be expected, which would be an acceptable agreement.  

\subsection{Tritium activity}
$^3$H activity was measured in all three targets. In addition, the graphite samples were sliced horizontally with regard to the beam direction, and the tritium was measured in each slice. In the 6 mm wide rim of the target, 3 to 5 slices could be made, each of which have different widths. In the simulation, 5 slices symmetrically around the center (s. figure \ref{geom}) with a width of 6/5 mm were modeled. In the simulation, a pencil beam is assumed whereas the real beam has a Gaussian distribution in vertical and horizontal direction with standard deviation of 0.5 to 1 mm (FWHM = 1.2 to 2.3 mm). Therefore, it is expected that the calculated distribution is smaller and particularly the peak maximum higher than the measured distribution. 

\begin{figure}[ht]
\begin{center}
\includegraphics[width=14cm]{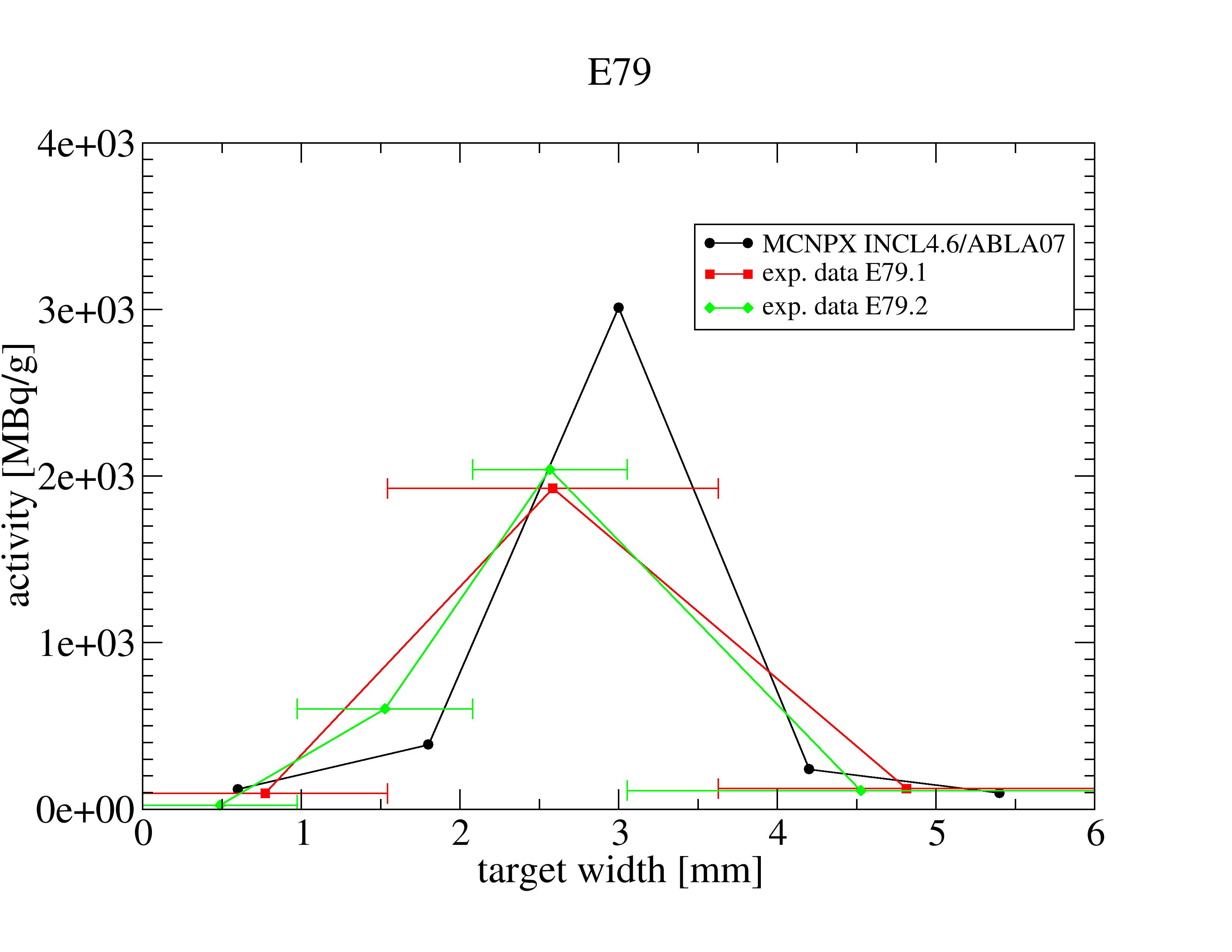}
\end{center}
\caption{\label{tritiumE79}Comparison of the simulated (using INCL4.6/ABLA07) and measured tritium activity as a function of the target width for E79. }
\end{figure}   

In figure \ref{tritiumE79}, the experimental results for the samples from E79 are compared to the MCNPX2.7.0 simulation using INCL4.6/ABLA07. Two samples were cut into 3 and 4 slices. The two measurements agree well to each other. The width of each slice is indicated by the horizontal error bar. Some slices are smaller, others larger. The results are not corrected  for bin sizes as then a shape of the distribution would need to be known or assumed. As expected, the experimental distribution is wider than the calculated one assuming a pencil beam. The experimental distributions seem to be slightly off center. The ratio of the Monte Carlo result and the measurement is 1.11, i.e. both are in very good agreement.   

\begin{figure}[hb]
\begin{center}
\includegraphics[width=14cm]{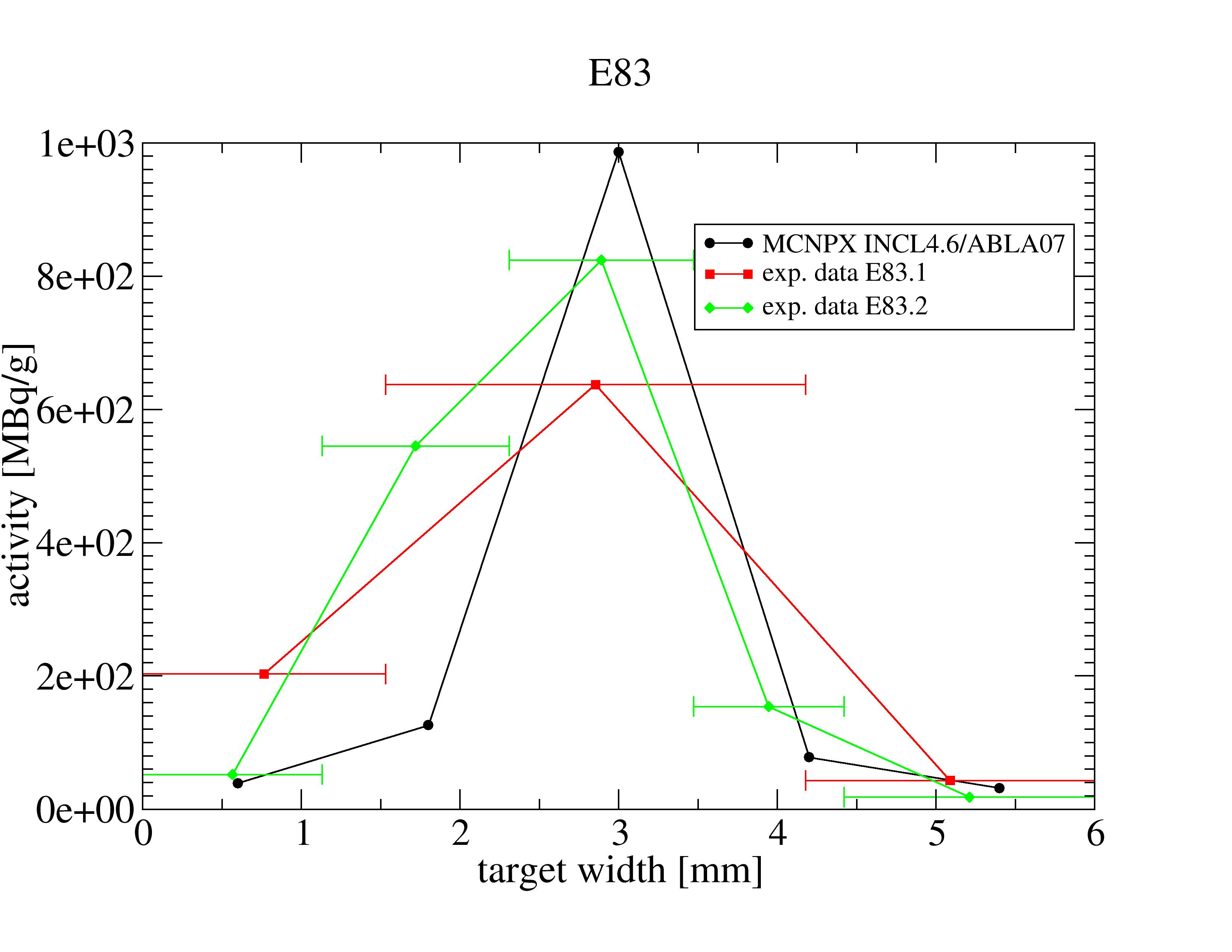}
\end{center}
\caption{\label{tritiumE83}Comparison of the simulated (using INCL4.6/ABLA07) and measured tritium activity as a function of the target width for E83. }
\end{figure}   

In figure \ref{tritiumE83}, the two measured tritium distributions in E83 are compared to the simulation. The experimental samples are cut in three and five slices and agree well to each other. The maximum is in the center whereas a bump is seen on the left. This might be due to an asymmetric beam distribution, or due to the superposition of different positions of the beam. The second case is more likely as the beam is indeed shifted to one or the other side depending which beam lines (left or right from Target E) has priority. The activity ratio for tritium from simulation to measurement is 0.79. Again, a reasonable agreement is achieved. 

\begin{figure}[ht]
\begin{center}
\includegraphics[width=14cm]{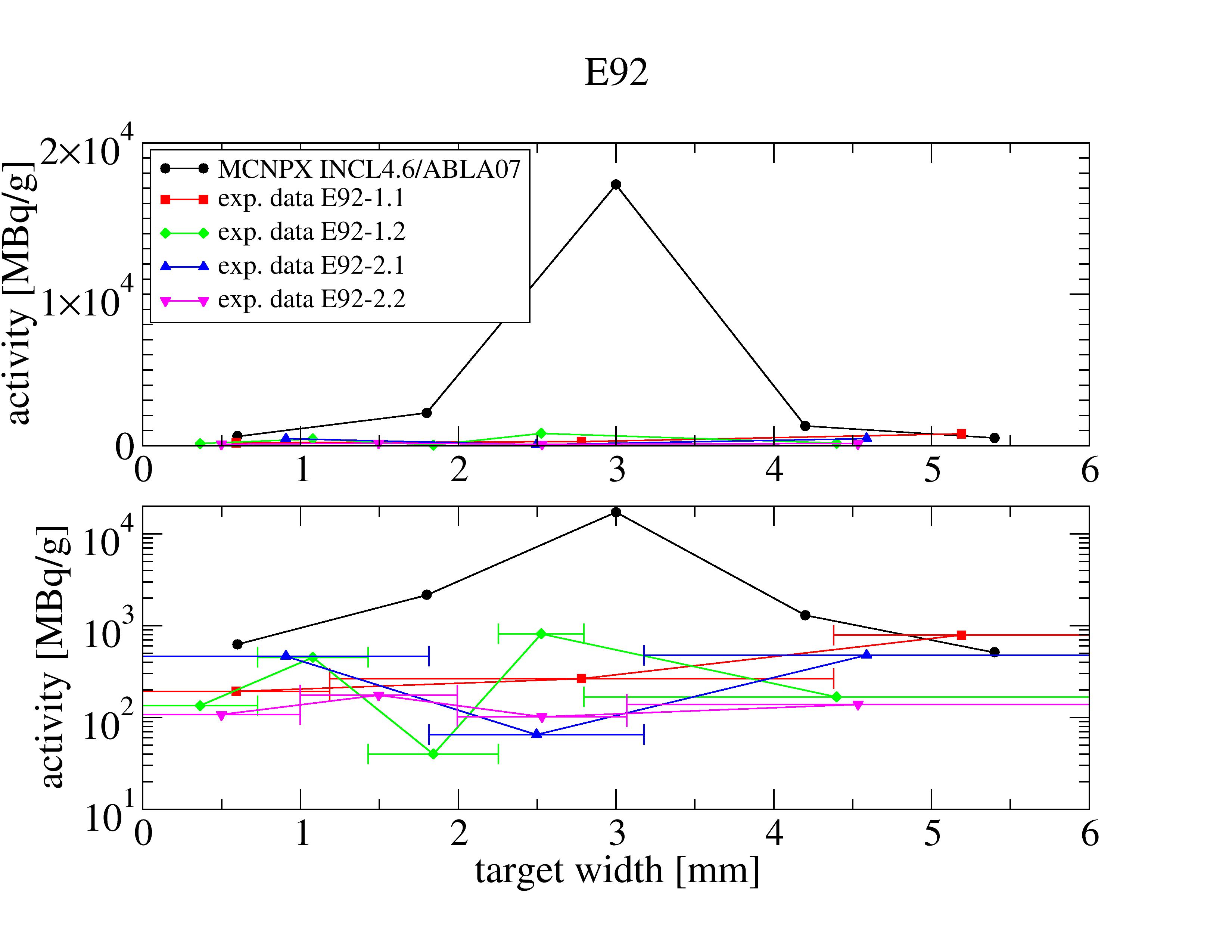}
\end{center}
\caption{\label{tritiumE92}Comparison of the simulated (using INCL4.6/ABLA07) and measured tritium activity as a function of the target width for E92. }
\end{figure} 

In figure \ref{tritiumE92}, the results for the samples from E92 are shown. Four samples were cut into three to five slices. However, none of the experimental distributions shows a clear peak. In addition, the measured tritium activity is much smaller than predicted from the calculation. The ratio of simulated and measured activity is 1/12, i.e. 90 \% of the tritium is missing. Using BERTINI-DRESNER-RAL instead of INCL4.6/ABLA07 would lead to 40 times less tritium.  In general, due to the high operation temperature ($\approx$ 1500 $^o$C) of the target, it is possible, that tritium already outgassed. However, the question is, why this happens in E92 only and not in E79 and E83.  E92 was operated at a bit higher current. However, since the target is cooled by radiation, the operation temperature is changed only slightly (about 50 to 100 $^o$C). However, in metals hydrogen starts to diffuse out at temperatures larger than  250 $^o$C. Another difference is that E92 is made from a different grade than E79 and E83. Although the E92 graphite grade R6510 is more radiation hard due to smaller grains,  the target E92 suffered from the largest beam charge, 4 times more than E79 and 16 times more than E83. It is known that particle radiation changes the structure of the material. The charge of about 29 Ah corresponds to 3 DPA (Displacements Per Atom). This means that every atom in the graphite structure is moved three times. Many defects in the structure occur, which facilitates the movement and outgassing of hydrogen. 


\subsection{Activities from impurities}
As already mentioned, the activities of the gamma emitters $^{22}$Na, $^{26}$Al, $^{44}$Ti and $^{54}$Mn were also detected. These radioisotopes are not produced by carbon, but from impurities with larger proton number like Na, Al, Ti and Mn. These elements were indeed detected in the material. However, first they vary with graphite grade, and second they lead to much larger activities than the carbon-born isotopes. Therefore, it was decided to apply another analysis method. For the mentioned elements, simulations were performed with pure carbon plus 10 ppm of one element as impurity. The resulting activity of the leading radioisotope (e.g. $^{22}$Na when using Na as impurity) was then compared to the measured activity. From this, the content of the impurity of the sample was deduced and compared to the abundance in the material analysis measured with ICP-OES. The investigations regarding the activities from impurities were performed for the samples from E92 and with the model INCL4.6/ABLA07 only. In table \ref{imp92} the second column contains the experimental data from E92 whereas the last column shows the results from the simulation using besides carbon 10 ppm of the element indicated in the third column. 

\begin{table}[ht]
\begin{center}
\caption{\label{imp92}Experimental and simulated activities for the radioisotopes from impurities as indicated.}
\begin{tabular}{lrlr}
\br
nuclide & exp. activity [Bq/g] & material comp. & MC activity [Bq/g]\\
\mr 
$^{22}$Na  & 52 & C + 10 ppm Na    & 1.52 10$^5$ \\
$^{26}$Al   & 0.3 & C + 10 ppm  Al    &  1.87 \\
$^{44}$Ti   & 16  &  C + 10 ppm Ti     &  1454 \\
$^{54}$Mn & 2.5 &  C + 10 ppm Mn   &  2.14 10$^4$ \\
 \br
\end{tabular}
\end{center}
\end{table}

As can be already seen from table \ref{imp92}, a small amount of Na or Mn leads to a large activity of $^{22}$Na and $^{54}$Mn, respectively . In addition, 10 ppm Al also produces 4.1 10$^4$ Bq/g $^{22}$Na and 10 ppm Mn leads to 957 Bq/g $^{44}$Ti. Therefore, the amount of the impurities estimated from the comparison of measured and simulated activities in table \ref{imp92} have to be seen as maximum values. These values are summarized in table \ref{ppm92} and are compared to the ICP-OES material analysis of R6510. 

Further, other impurities also lead to sometimes significant contributions to the radioisotopes listed in table \ref{ppm92}. For this a simulation was performed adding 10 ppm Fe to pure Carbon. As expected, a significant amount of $^{54}$Mn (9500 Bq/g) and $^{44}$Ti (1300 Bq/g) are produced. Comparing to table \ref{imp92}, this is half of the $^{54}$Mn activity obtained with 10 ppm Mn and almost the same amount of $^{44}$Ti as produced with 10 ppm Ti. 

\begin{table}[ht]
\begin{center}
\caption{\label{ppm92}Content of trace elements in the first column deduced from the comparison of calculational  and experimental activity of the radionuclides in the second column as well as compared to the ICP-OES measurement.}
\begin{tabular}{llrrr}
\br
element & nuclide & (MC/exp.) [ppm] & ICP-OES [ppm] & ICP-OES/MC \\
\mr 
Na & $^{22}$Na  & 0.0034 & 13 & 3805\\
Al & $^{26}$Al   & 1.61     & 5.8 & 4\\
Ti & $^{44}$Ti   & 0.11      & 0.66  & 6\\
Mn & $^{54}$Mn & 0.0012 & 0.12 & 103\\
\br
\end{tabular}
\end{center}
\end{table}

The third column of table \ref{ppm92} was calculated by comparing the measured and simulated activities in table \ref{imp92}. In the Monte Carlo simulation, 10 ppm of the trace element was always used. From the measured $^{22}$Na activity just 0.0034 ppm should be present in the sample while 13 ppm Na were measured using ICP-OES. However, Na is present everywhere as an impurity and therefore it is likely that the large amount measured in the sample is a contamination during the measurement. The measured amount of 0.12 ppm Mn is already quite small. However, the concentration deduced with the help of the simulation is even a factor 100 smaller. This very small amount would not be measurable by ICP-OES. In general, determining impurities in the sub-ppm region is very difficult and requires another method like ICP-MS (Induced Coupled Plasma Mass Spectrometry) as well as very clean conditions. Therefore, it can be concluded that the graphite used for target E is much purer than assumed from the material analysis.  This also explains the observation that after a few years of cooling the calculated dose rates of all former graphite targets were much higher than measured. When the $^{7}$Be is decayed, the dose rate after a few years is mainly determined by $^{22}$Na, which is according to table \ref{ppm92} about 3 orders too high in the formerly assumed material composition. 

\section{Summary results and conclusions} \label{sum}
The results for the comparison of measured and calculated activities of the graphite target E are summarized in the tables \ref{sum79}, \ref{sum83}, \ref{sum92_I} and \ref{sum92_B}. 

\begin{table}[ht]
\begin{center}
\caption{\label{sum79}Summary of the results for measured and calculated activities on samples from E79. The simulation was performed with MCNPX2.7.0 using INCL4.6/ABLA07.}
\begin{tabular}{lrrr}
\br
nuclide & exp. activity [Bq/g] & MC activity [Bq/g] & MC/exp.\\
\mr 
$^3$H & 6.94E+08 & 7.73E+08 & 1.11\\
$^{10}$Be & 4004 & 5943 & 1.48\\
$^{14}$C & 2041 & 1022 & 0.50\\
\br
\end{tabular}
\end{center}
\end{table}

\begin{table}[ht]
\begin{center}
\caption{\label{sum83}Summary of the results for measured and calculated activities on samples from E83. The simulation was performed with MCNPX2.7.0 using INCL4.6/ABLA07.}
\begin{tabular}{lrrr}
\br
nuclide & exp. activity [Bq/g] & MC activity [Bq/g] & MC/exp.\\
\mr 
$^3$H & 3.19E+08 & 2.52E+08 & 0.79\\
$^{10}$Be & 546 & 1650 & 3.02 \\
$^{14}$C & 244 & 284 & 1.16\\
\br
\end{tabular}
\end{center}
\end{table}

\begin{table}[ht]
\begin{center}
\caption{\label{sum92_I}Summary of the results for measured and calculated activities on samples from E92. The simulation was performed with MCNPX2.7.0 using  INCL4.6/ABLA07.}
\begin{tabular}{lrrrrr}
\br
nuclide & exp. activity [Bq/g] & MC activity [Bq/g]  & MC/exp.\\
\mr 
$^3$H & 3.60E+08 &  4.43E+09 & 12.30\\
$^{7}$Be & 594 & 1092 & 1.84\\
$^{10}$Be & 13567  & 26300 & 1.94\\
$^{14}$C & 4648  & 4270 & 0.92\\
\br
\end{tabular}
\end{center}
\end{table}

\begin{table}[ht]
\begin{center}
\caption{\label{sum92_B}Summary of the results for measured and calculated activities on samples from E92. The simulation was performed with MCNPX2.7.0 using BERTINI-DRESNER-RAL.}
\begin{tabular}{lrrrrr}
\br
nuclide & exp. activity [Bq/g] & MC activity [Bq/g]  & MC/exp.\\
\mr 
$^3$H & 3.60E+08 & 1.45E+10	 & 40.17\\
$^{7}$Be & 594 & 784 & 1.32\\
$^{10}$Be & 13567 & 13680 & 1.01\\
$^{14}$C & 4648 & 64.92 & 0.01\\
\br
\end{tabular}
\end{center}
\end{table}

The uncertainty for the $^{14}$C measurement is 15 \% and its detection limit is 500 Bq/g, meaning the $^{14}$C activity of E83 measurement is below this limit. However, all measurements at the three targets regarding $^{14}$C are for the first time well reproduced by using the cross section model INCL4.6/ABLA07 in MCNPX2.7.0. The default model, BERTINI-DRESNER-RAL used in former analyses underpredicts the $^{14}$C activity by two orders of magnitude. The main reaction responsible for the large $^{14}$C production is neither the neutron capture in $^{13}$C nor a contamination with nitrogen. Reactions with tritons and deuterons on $^{12}$C lead to the additional $^{14}$C activity. This indirect mechanism required 10$^8$ primary protons to be simulated in order to achieve good result statistics. 

From the results of E92, it seems that $^3$H outgassed during operation of the target. This is supported by the flat distribution of $^3$H along the target width (horizontal direction), whereas in E79 and E83, the peak of the tritium activity is clearly visible around the center of the target width. Candidates for an explanation are the different graphite grade or radiation damage. 

Comparing the activities from the simulations using BERTINI-DRESNER-RAL and INCL4.6/ABLA07 for the Be-isotopes, $^{7}$Be and $^{10}$Be, the default model of MCNPX2.7.0 fits the experimental data better. INCL4.6/ABAL07 is higher than the measurement for all cases, and about 40 \% higher for $^{7}$Be activity and 90 \% for $^{10}$Be activity than the simulation using BERTINI-DRESNER-RAL. 

The analysis of the activities from impurities revealed a much purer graphite than assumed from the ICP-OES material analysis. The impurities seem to be in the sub-ppm region, which might require another measurement method and procedures to avoid contamination. Before the investigation in this report, it was not expected that impurities on the sub-ppm level would lead to a measurable activity. Not measured, but discussed in the introduction, is the activity of $^{36}$Cl, which is thought to be a problematic nuclide due to its long life time and its ability to migrate out of the component. In the EK94 grade graphite, 1.1 ppm Cl was determined. Using MCNPX2.7.0 and INCL4.6/ABLA07 this would lead to 0.12 Bq/g  $^{36}$Cl. The activity is smaller than expected due to the fact that $^{36}$Cl is produced by capturing thermal neutrons. This is due to the fact that few thermal neutrons are produced in the meson target compared to a nuclear reactor.   

\section*{Acknowledgement}
We would like to thank Tanja Wieseler und Sabrina L\"uthi for the chemical preparation of the graphite samples and the measurements of the activities. Further, we are grateful to Dr. Jean-Christophe Davide for giving us access to INCL4.6/ABLA07 and helping us with the implementation into MCNPX2.7.0.

\section*{References}
\bibliography{ARIA17}

\end{document}